\documentclass[letterpaper,twocolumn,10pt]{article}
\usepackage{usenix-2020-09}

\usepackage{tikz}
\usepackage{amsmath}
\usepackage{filecontents}
\usepackage{xspace}
\usepackage[normalem]{ulem}
\usepackage{graphicx}
\usepackage{subcaption}
\usepackage{caption}
\usepackage{graphicx}
\usepackage[most]{tcolorbox}
\usepackage{booktabs}
\usepackage{enumitem}
\usepackage{amssymb}

\newtcolorbox{takeaway}{
  colback=gray!15,    
  colframe=gray!50,   
  boxrule=0.5pt,      
  arc=3pt,            
  left=4pt, right=3pt, top=3pt, bottom=1pt
}

\definecolor{darkred}{RGB}{180,0,0}
\definecolor{darkblue}{RGB}{0,0,160}

\usepackage{xcolor}

\begin{document}

\date{}

\title{\LARGE \bf Trinity: \Large Efficient GPU-based VectorStore-Prefill-Decode Disaggregation for LLM Serving}

\title{\LARGE \bf Trinity: \Large Disaggregating Vector Search for Prefill-Decode \\Disaggregation in LLM Serving}


\author{
{\rm Yi Liu, Chen Qian} \\
 \it University of California Santa Cruz
}

\newcommand{\sys}{\textbf{Trinity}\xspace}

\maketitle

\begin{abstract}

Prefill and decode (PD) disaggregation separates prompt prefill and token-by-token decode stages into distinct GPU pools and has become the dominant architecture for large-scale LLM serving in industry. 
Also, retrieval tasks via vector search remains entangled with the model inference process, like heterogeneous RAG requests and prompt answer caches, inflating tail latency. 
We are motivated to investigate how vector search should be orchestrated along with PD disaggregation with a dedicated deployment architecture without violating SLOs in various retrieval workloads.
We present \sys, a practical framework that consolidates all retrieval into a single, shared vector-search GPU pool and make it work with PD disaggregated LLM serving in match. \sys introduces (1) a novel architecture for deploying GPU-based vector search service in PD disaggregation. (2) Continuous batching for vector search that make full used of GPUs under heterogeneous queries; (3) Stage-aware scheduling that preempts vector search requests between both decode and prefill tasks.
Our analysis demonstrates that an independent vector search pool can serve diverse vector retrievals while sustaining high throughput and low tail latency for LLM serving with PD disaggregation.

\end{abstract}

\vspace{-1.5ex}
\section{Introduction}
\vspace{-.5ex}
Retrieval-augmented generation (RAG) mitigates hallucinations and enhances output quality by retrieving semantically relevant document chunks during inference~\cite{piperag,tigervector,rago,ragdoll,ragcache,flexgen,co,langchain}. Vector search~\cite{milvus,diskann,distributed-faiss,distributedann} serves as the foundation of RAG services and prompt cache (e.g., GPTCache~\cite{gptcache}), enabling efficient nearest neighbor search from large-scale embeddings. 
Since exact kNN search is computationally prohibitive at scale, production systems adopt approximate nearest-neighbor (ANN) indexes, such as IVF/IMI with PQ or OPQ compression, or graph-based structures like HNSW, to trade slight recall loss for substantial gains in throughput and memory efficiency. 
To further accelerate large-scale vector retrieval, recent GPU-based ANN schemes~\cite{zhao2020song,cagra,pilotann,tian2025towards,co} are exploited to batch distance computations, leverage high-bandwidth HBM, and overlap data transfer, search, and re-ranking. These optimizations enable sub-millisecond query latency even under high-throughput workloads. 
For example, CAGRA~\cite{cagra}, a GPU-optimized graph ANN implementation in cuVS, adopts fixed-degree graphs and warp-efficient traversal to sustain high recall while maintaining low tail latency with batched execution.

\begin{figure}[!t]
  \centering
    \includegraphics[width=.85\linewidth]{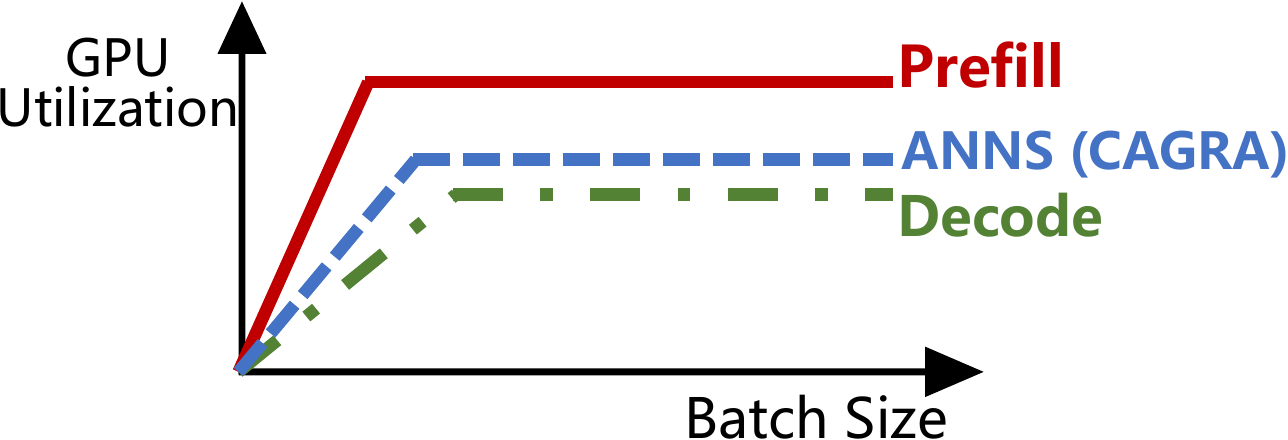}
  \caption{Roofline of Vector Search, Prefill and Decode in LLM Serving.}
  \label{fig:roofline}
  \vspace{-2.ex}
\end{figure}

\begin{figure*}[!t]
  \centering
  \begin{subfigure}{0.45\textwidth}
    \captionsetup{labelformat=empty}
    \includegraphics[width=\linewidth]{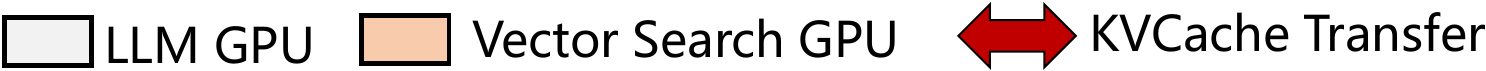}
  \end{subfigure}\\
  \vspace{2ex}
  \setcounter{subfigure}{0}
  \captionsetup{labelformat=parens,labelsep=space}
  \begin{subfigure}{0.25\textwidth}
    \includegraphics[width=\linewidth]{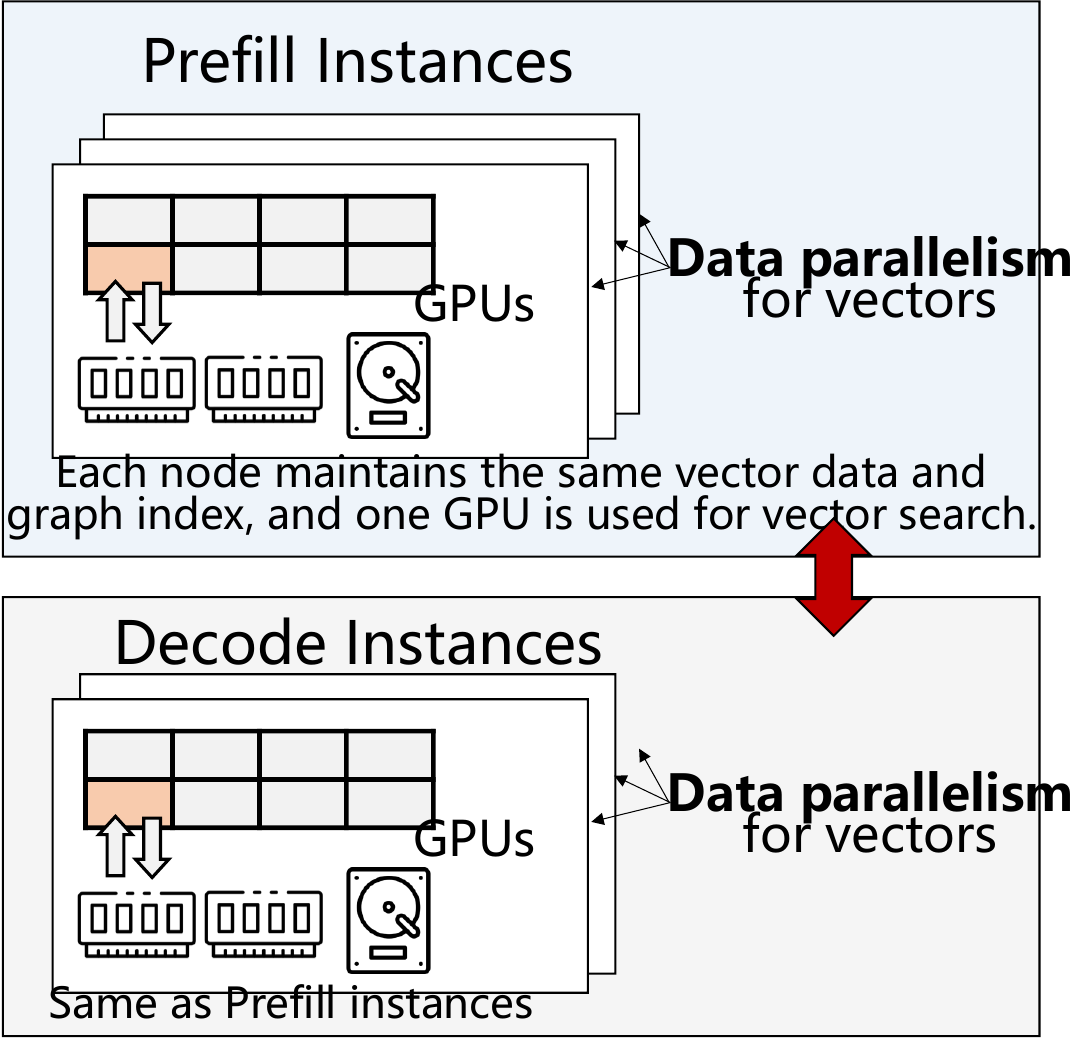}
    \caption{Vector search GPU is coupled with LLM GPU in the same servers.}
    \label{fig:arch:0}
  \end{subfigure}
  \hspace{3ex}
  \begin{subfigure}{0.345\textwidth}
    \includegraphics[width=\linewidth]{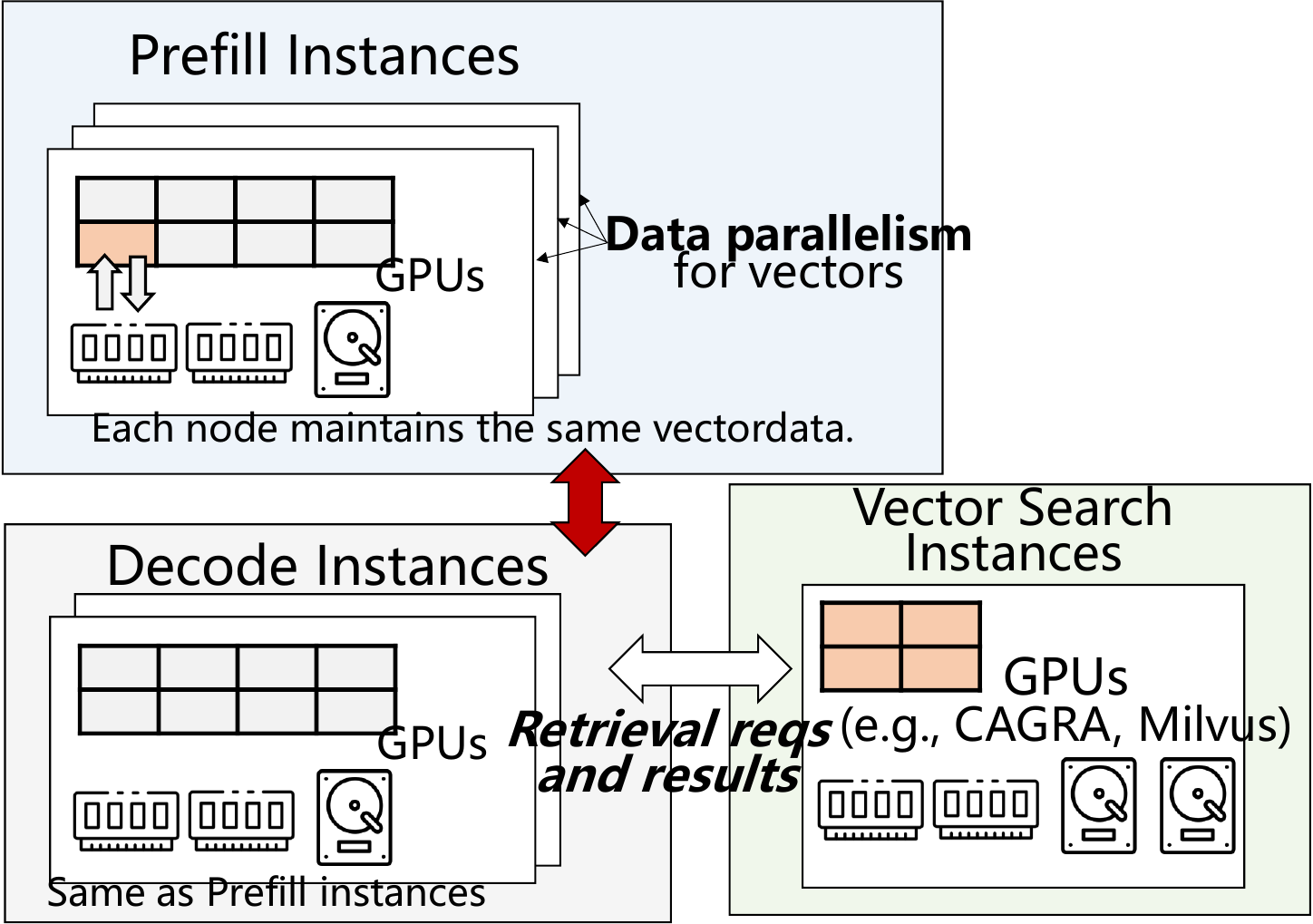}
    \caption{Vector search GPU is decoupled from Decode instances, but being together with Prefill Instance.}
    \label{fig:arch:1}
  \end{subfigure}
  \hfill
  \begin{subfigure}{0.34\textwidth}
    \includegraphics[width=\linewidth]{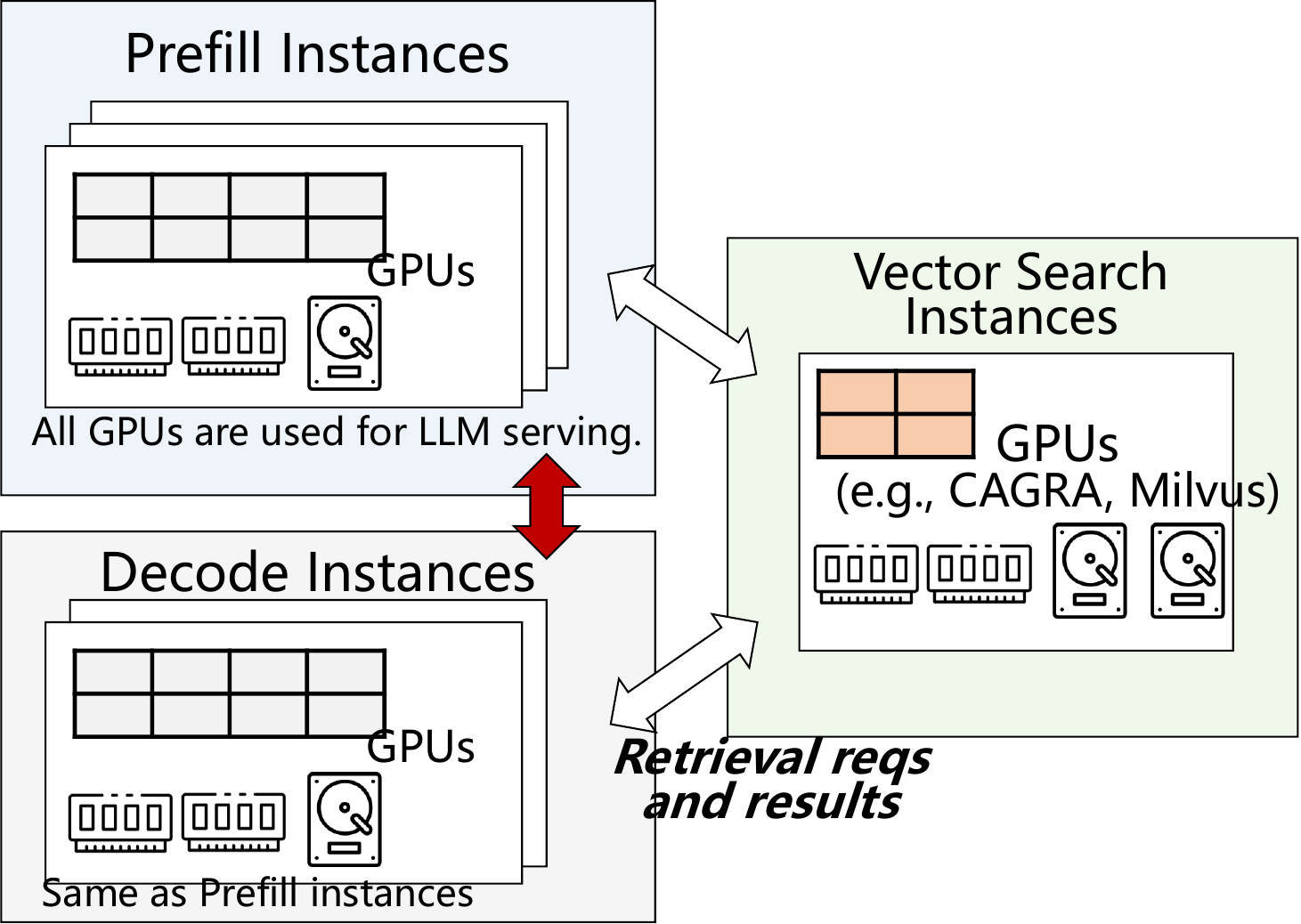}
    \caption{Vector search is offloaded to an independent GPU pool.}
    \label{fig:arch:2}
  \end{subfigure}    
  \caption*{Figure 2. Architectures of serving GPU-based vector search instances in PD disaggregation.}
  \label{fig:arch}
  \vspace{-2.ex}
\end{figure*}

PD disaggregation~\cite{distserve,pdserve,sglang,vllm,ernie-4.5,chameleon,liu2025elasticmm} has emerged as a state-of-the-art paradigm for LLM serving in industry. Traditional monolithic deployments execute both stages on the same GPU, resulting in inefficient resource utilization since prefill and decode exhibit fundamentally different characteristics: the \textbf{prefill} stage is compute-intensive and benefits from high parallelism across long input sequences, while the \textbf{decode} stage is memory-bound for generating tokens sequentially with billion-scale parameters of experts~\cite{ernie-4.5,deepseek-v3,deepseek-r1} and MLP. By decoupling these stages into distinct GPU or node pools, PD disaggregation enables independent scaling, higher hardware utilization, and lower latency, and it has become central to modern LLM infrastructures with low cost.

While numerous studies~\cite{chameleon,cagra,ragdoll,rago} have explored how to integrate vector search into existing LLM inference, the interaction between GPU-based vector retrieval and current PD disaggregation remains largely underexplored. 
In a disaggregated serving architecture, the prefill stage typically performs a single retrieval to initialize context at the beginning, whereas the decode stage may require multiple rounds of retrieval as generation progresses to maintain topical relevance or inject new knowledge. This asymmetry introduces unique challenges in coordinating retrieval frequency, managing cross-stage communication, and optimizing end-to-end latency across disaggregated resources.

In this work, we explore several GPU disaggregation architectures for incorporating vector search service into large-scale LLM serving systems and analyze how different designs coordinate these GPUs efficiently. Building on these insights, we design \sys, a disaggregated serving framework with three key innovations.
(1) To reduce document retrieval latency, we explore and analysis three potential architectures for running vector search either coupled with P/D GPUs or decoupled into shared GPU pools under PD disaggregation.
(2) To improve GPU utilization for vector search, we introduce a continuous batching mechanism that executes graph-based vector search efficiently on GPUs with a dynamically running batch of concurrent queries.
(3) To guarantee SLOs for both prefill and decode instances (e.g., time-to-first-token (TTFT)), \sys employs an adaptive scheduling and preemption scheme using multiple priority queues to balance latency and throughput across serving stages.

\section{Motivation}
In this section, we contrast the computation and storage need for vector search, prefill, and decode, and show that each has a different optimal batch size. 
We observe that both decode and vector search are primarily \emph{memory-bound}: greedy graph traversal repeatedly loads neighbor vectors and indices at each traversal step, resulting in low Arithmetic Intensity (AI). 
Then, we analyze the GPU utilization roofline of each stage to identify their plateaus and the batch sizes at which they saturate, as shown in Fig.~\ref{fig:roofline}.

Let $u(\cdot)\in[0,1]$ be GPU utilization.
Let $\mathcal{P}_{\text{peak}}$ be peak compute (FLOP/s) for the chosen datatype,
$\mathcal{B}_{\text{mem}}$ the effective memory bandwidth (B/s),
and $\text{AI}$ the arithmetic intensity (FLOP/Byte).
Let $B$ be batch size (sequences) for LLM. For ANN search let $Q$ be the number of concurrent queries.
Define the generic roofline as:
\[
u_{\max}\;\triangleq\;\min\!\Big(1,\; \frac{\text{AI}\cdot \mathcal{B}_{\text{mem}}}{\mathcal{P}_{\text{peak}}}\Big).
\]
We model the pre-plateau rise with a saturation scale $X_{\text{sat}}$ and an optional sublinearity exponent $\alpha\!\in\!(0,1]$:
\[
u(X)\;\approx\;\min\!\Big(u_{\max},\; \big(\tfrac{X}{X_{\text{sat}}}\big)^{\alpha}\Big),
\quad X\in\{B,Q\}.
\]
The pre-plateau slope at small $X$ is $\frac{\mathrm{d}u}{\mathrm{d}X}\big|_{X\ll X_{\text{sat}}}\!\approx\!\alpha/X_{\text{sat}}$.

\paragraph{(1) Prefill.} 
Large GEMMs operations means high $\text{AI}$, which indicates it is compute-bound.
\[
u_{\text{prefill}}(B)\;\approx\;\min\!\Big(1,\; \big(\tfrac{B}{B_{\text{sat,p}}}\big)^{\alpha_p}\Big),
\qquad
u^{\max}_{\text{prefill}}\approx 1.
\]
$B_{\text{sat,p}}$ is the smallest batch size at which the prefill phase
(big GEMMs) effectively saturates the GPU. Beyond this point, increasing $B$
yields negligible utilization gains. Thus, its plateau can reach $\approx 100\%$ once tensor cores are filled.

\paragraph{(2) Decode.}
One new token per request, multi-layer KVCache reads will be incurred means it is low $\text{AI}$ (memory-bound).
\[
u_{\text{decode}}(B)\;\approx\;\min\!\Big(u^{\max}_{\text{decode}},\; \big(\tfrac{B}{B_{\text{sat,d}}}\big)^{\alpha_d}\Big),
\quad
u^{\max}_{\text{decode}}\approx \frac{\text{AI}_{\text{dec}}\cdot \mathcal{B}_{\text{mem}}}{\mathcal{P}_{\text{peak}}}.
\]
\emph{Plateau:} flat, bandwidth-limited (often well below $100\%$).

\paragraph{(3) GPU ANN (e.g., CAGRA in cuVS).}
This phase is dominated by graph traversal and distance computations, because per-vector bytes dominate, AI is low. Thus, it is memory-bound, even with only a small compute-leaning re-rank stage when queries are batched.

\[
u_{\text{cagra}}(Q)\;\approx\;\min\!\Big(u^{\max}_{\text{cagra}},\; \big(\tfrac{Q}{Q_{\text{sat}}}\big)^{\alpha_a}\Big),
\quad
u^{\max}_{\text{cagra}}\approx \frac{\text{AI}_{\text{graph}}\cdot \mathcal{B}_{\text{mem}}}{\mathcal{P}_{\text{peak}}}.
\]
\emph{Plateau:} bandwidth roof, which is similar order to decode at equal precision.

\section{Methodology}
\vspace{-1.5ex}
\subsection{Architecture}
\vspace{-.5ex}
From the motivation, we know that vector search and the prefill/decode phases have different optimal batch sizes, suggesting they should run on separate GPUs. 
To reduce the retrieval latency, the most intuitive way is co-locate a vector search GPU on the P/D servers, thus, fast intra-node links (e.g., NVLink) can reduce the data transfer latency for retrieval. 
The trade-off is that P/D server has to reserve a GPU for vector search, which reduces available computation capacity for LLM inference and can introduce bandwidth contention. 
In the following, we discuss co-location by trading off SLO improvements on vector search against lost inference throughput and contention to see when co-location is worthy.

\noindent\textbf{Coupled placement of vector search and LLM GPUs.}
Fig.~\ref{fig:arch:0} shows a design where each prefill and decode server co-locates one vector-search GPU with the LLM GPUs. The vector database
(embeddings and graph index) is replicated per node and sharded across the local vector GPU via Data Parallelism (DP). At query time, both prefill and decode workers send retrievals to the in-node vector GPU over NVLink, minimizing retrieval latency and avoiding network hops. 
Also, the per-node replication of the graph/index inflates memory footprint and complicates updates, we can also make several LLM nodes to share one vector search GPU.

However, in the decode stage we typically use expert parallelism (EP): each GPU stores parameters for a subset of MLP experts and participates in dispatch/combine kernels. If one GPU on a node is reserved for vector search, an additional GPU must be provisioned elsewhere to host the displaced experts, pushing part of the EP traffic inter-node. The extra
network latency during dispatch/combine can outweigh the latency saved by in-node retrieval. 

\noindent\textbf{Decoupled vector search from Decode but co-located with Prefill.}
The Fig.~\ref{fig:arch:1} depicts a layout where vector search GPUs are running in a separated in a independent pool only for serving retrieval requests from decode instances, but some of them are co-located with the prefill servers with DP.
Prefill sends retrieval requests to the in-node vector GPUs over NVLink/PCIe, while decode reaches the vector pool over the network.

Co-locating vector search with prefill removes bandwidth contention between decode and vector search , and gives prefill the shortest retrieval path with intra-node fabric, reducing retrieval latency for prefill instances.
However, even though prefill is compute-bound, tensor parallelism (TP) introduces collective synchronization that sits on the critical path. 
In particular, the Q/K/V projections and subsequent attention blocks require all-reduces to merge partial GEMM results across TP shards. 
The saved microseconds from in-node document retrieval are typically smaller than the added or unchanged collective latency for merging these GEMMs, especially at higher TP degrees. Consequently, the fast retrieval path cannot compensate for the step-time dominated by TP collectives, so this architecture gains are limited despite co-location with prefill.

\noindent\textbf{Offloading vector search to an independent GPU pool.}
Based on the preceding analysis of rooflines and saturation scales, we derive the design shown in Fig.~\ref{fig:arch:2}, vector search is offloaded to a separate GPU pool, while all GPUs on prefill and decode servers are dedicated to LLM serving. Retrieval requests and results traverse the network between the LLM tiers and the vector-search tier.

The vector-search cluster shards embeddings and the graph index across its own GPUs (e.g., CAGRA, Milvus). Prefill and decode instances issue retrieval RPCs to this pool and receive top-$k$ document chunks. No GPU on the LLM side is reserved for vector search, thus, also eliminating intra-node and inter-node bandwidth contention for both prefill and decode.

\vspace{-2ex}
\subsection{Continuous batching on vector search.}
\label{sec:continuous}

\begin{figure}[!t]
    \hspace{-3.ex}
    \includegraphics[width=.5\textwidth]{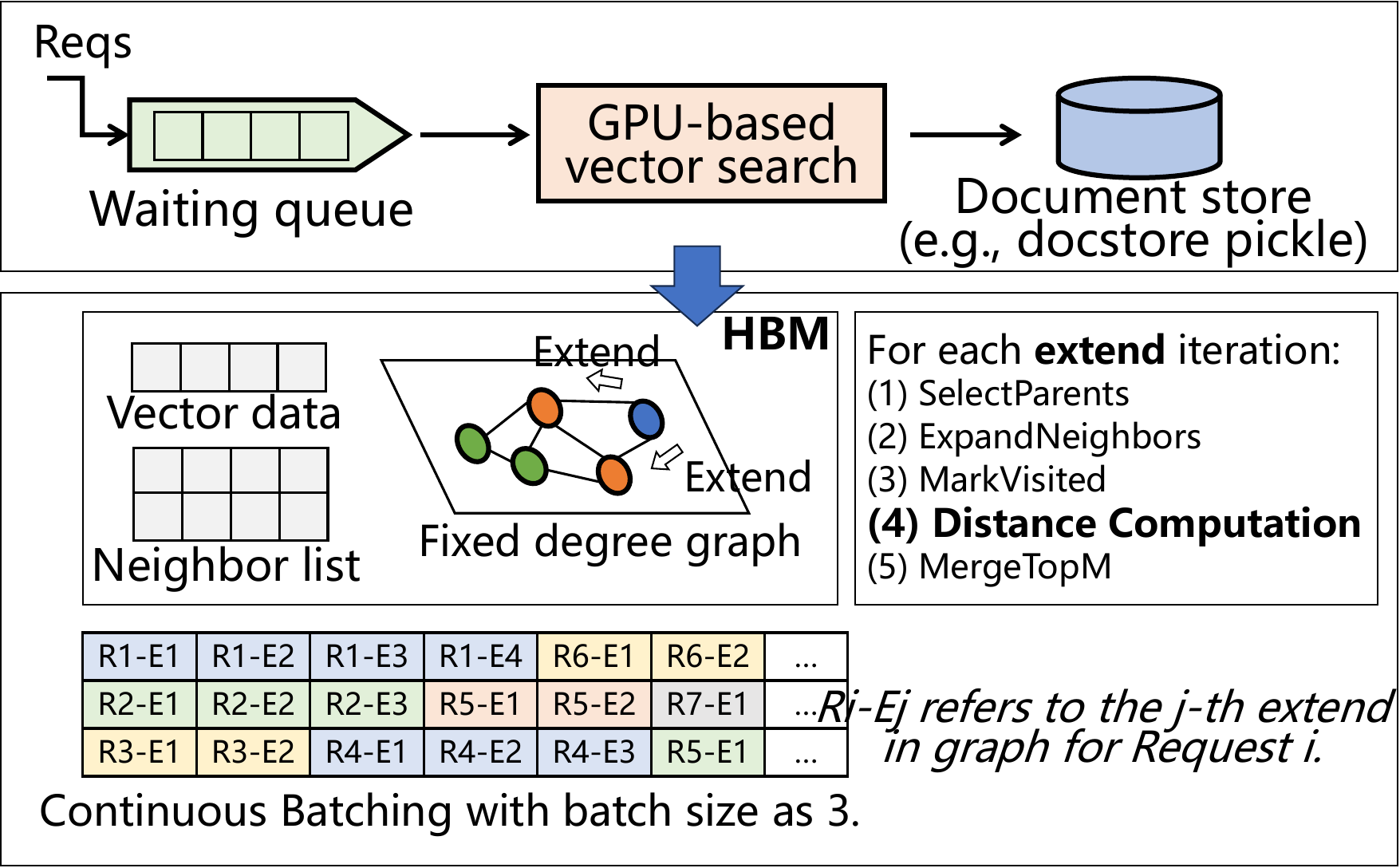}
    \caption{Continuous batching for vector search in \sys.}
    \label{fig:continuous}
    \vspace{-2.5ex}
\end{figure}

\noindent To improve GPU utilization and solve latency jitter that arise when CAGRA executes each request in batches, we design a continuous-batch execution that treats one \emph{extend} step on graph as the unit and interleaves many requests precisely at that granularity. 
As illustrated in Fig.~\ref{fig:continuous}, the GPU keeps the database vectors $\mathbf{X}\!\in\!\mathbb{R}^{N\times d}$ resident in HBM together with a fixed-degree neighbor list $G\!\in\!\mathbb{N}^{N\times D}$, so distance and expansion are pure device-memory operations. 
Each request maintains compact device-side state: an internal $\texttt{topM}$ (ids and distances) used for storing candidates by far, a \texttt{visited} table used to admit only first-seen candidates to computation. 
A scheduling loop advances all active requests through one \textbf{\emph{extend}}: for each request we select up to $p$ non-expanded parents from its $\texttt{topM}$, read $D$ neighbors per parent from the neighbor list, filter by \texttt{visited}, and emit the surviving ids as \emph{distance tasks} into a global, cross-request task array. 
We then evaluate all accumulated distance tasks with a single fixed-shape kernel that uses warp-splitting teams with high SM occupancy and cuda graph. 
If the available tasks are fewer than the captured kernel size, we round up with masked dummies to preserve a stable operator shape. 
Then, the computed $(\text{id},\text{dist})$ are scattered back to their origin requests, locally merged with $\texttt{topM}$, and the consumed parents are marked \texttt{expanded}. Early-stop is decided independently per request when the candidates in $\texttt{topM}$ list are not changed, so completed requests exit immediately and their device buffers are recycled, while new arrivals are admitted and begin contributing tasks to the next distance batch without idling the GPU.

Compared to the original per-request batching, continuous batching yields a distinct set of advantages in a single, compatible design. 
First, cross-request aggregation converts many tiny, shape-varying distance evaluations into a single fixed-shape operator that maintains near-peak HBM throughput and SM utilization even when individual requests are short, uneven, or bursty. 
Second, using a small number of fixed-shape launches (or a captured CUDA Graph) amortizes CPU/driver overhead and reduces variance, producing smoother P50/P95 and predictable capacity at scale. 
Third, the kernel's stable shape with fixed number of degree $D$ and fixed captured batch size with safe padding simplifies memory planning.
All finished queries vacate resources immediately, and newcomers join the very next distance batch with only a short, configurable flush timeout bounding additional wait. 
Since CAGRA maintains all vectors and the fixed-degree graph in HBM, our design keeps vector search results accuracy/recall behavior intact while turning the global distance stage into a steady, high-throughput engine that better matches GPU hardware characteristics.

\subsection{Latency-aware requests scheduling for prefill/decode instances.}
\begin{figure}[!t]
    \hspace{-1.5ex}
    \includegraphics[width=.5\textwidth]{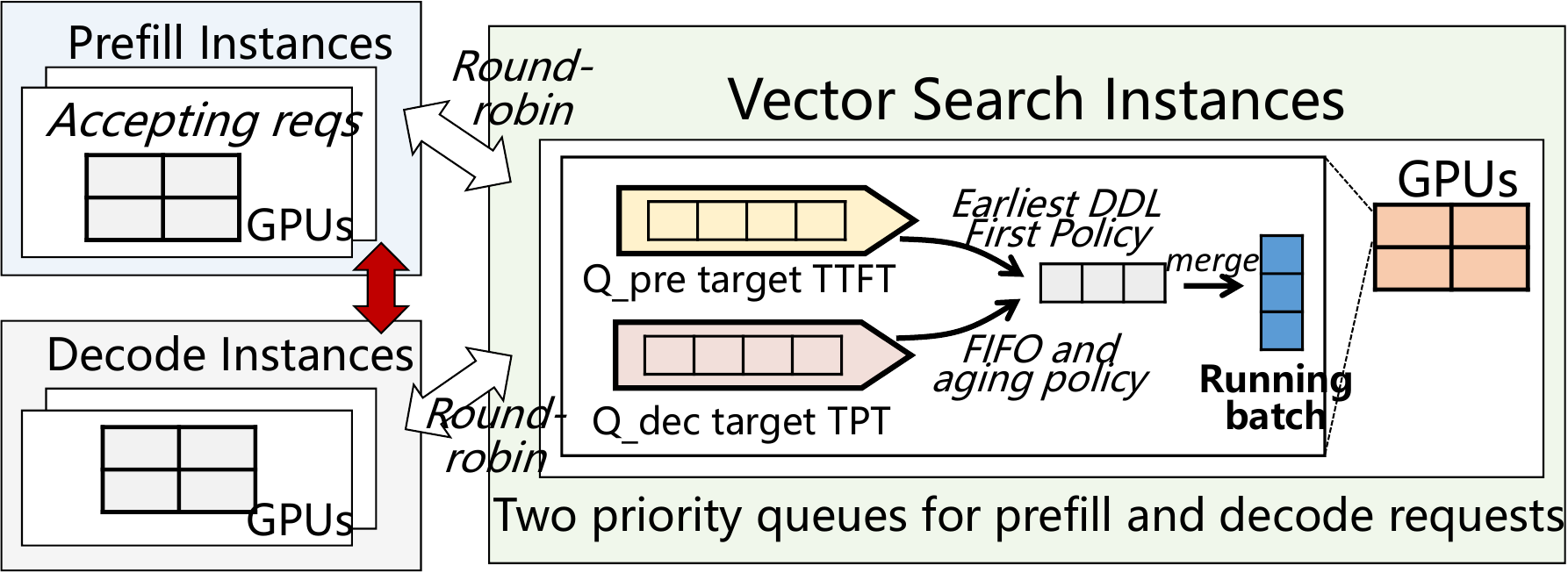}
    \caption{Priority queue-based scheduling for prefill/decode instances in \sys.}
    \label{fig:queue}
    \vspace{-2.5ex}
\end{figure}

To meet TTFT for prefill and TPT for decode with best efforts, we add a scheduler in the vector pool that coordinates search requests from both prefill and decode stages.
For prefill instances, it will make latency-critical RAG requests to retrieve prompt-relevant context upon it receives batched requests,, whereas decode emits periodic RAG probes every $\Delta$ tokens\footnote{There are many of different RAG search triggering policies, here we use a fixed-tokens policy as an example.} to preserve perplexity and relevance during new token generation.
A naïve policy that always favors vector search requests from prefill side will cause decode to wait on retrieval, it will leave token generation idling for context on decode side, thereby reducing average token throughput of ranks even if TTFT appears healthy. 
Conversely, always favoring decode will delay prefill stage, push TTFT beyond the latency budget, and fails to fill the KVCache transfer link (e.g., Mooncake) bandwidth between prefill and decode pools, enough to keep decode workers busy, and the PD disaggregation pipeline never saturates.
Thus, we design an adaptive scheduler that works with two requests queues, and feed the running batch with vector requests from prefill and decode sides flexibly, so that latency SLOs will be met and the GPU utilization achieved by PD disaggregation will be preserved.

As shown in Fig.~\ref{fig:queue}, we use two priority queues to schedule vector search requests: a high priority prefill queue $Q_{\text{pre}}$ and a lower priority decode queue $Q_{\text{dec}}$. 
When we select request candidates from two queues into the current running batch, we build a request buffer of length $N$, the scheduler sets $N$ to match the number of available slots in the current batch, reserves a fraction $r \in [r_{\min}, r_{\max}]$ for $Q_{\text{pre}}$ so that at least $rN$ entries come from prefill, and fills the remainder with $Q_{\text{dec}}$. 
If $Q_{\text{pre}}$ cannot supply enough entries at that moment, its unused share is immediately given to $Q_{\text{dec}}$. If $n_{\text{pre}}+n_{\text{dec}}<N$, the builder pads with masked dummy tasks to preserve the kernel’s fixed shape. 

For $Q_{\text{pre}}$ we apply Earliest Deadline First (EDF) policy and slack-driven scheduling approach: each request carries a deadline as $\textit{ddl}=t_{\text{arr}}+L_{\text{pre,max}}$ and an estimate of remaining extends $\tilde E$ in vector search, and we rank by $\textit{slack}=\textit{ddl}-(t_{\text{now}}+\tilde E \cdot T_{\text{ext}})$, where $T_{\text{ext}}$ is the measured average latency per extend. 
Furthermore, we assign $Q_{\text{pre}}$ a short flush timeout $\tau_{\text{pre}}$, ensuring urgent prefill search request can be appended into running batch without waiting for the global timeout.

For $Q_{\text{dec}}$ we use FIFO to select request candidates.  
Together with the prefill reservation $r$ and the short prefill timeout $\tau_{\text{pre}}$, FIFO allows decode to steadily consume the remaining capacity while prefill meets TTFT and keeps the KV cache pipeline saturated.

The scheduler materializes a pair $(n_{\text{pre}}, n_{\text{dec}})$ with $n_{\text{pre}}\ge rN$ and $n_{\text{pre}}+n_{\text{dec}}=N$, launches one fixed shape distance kernel with warp teams, and scatters results back for per request \texttt{topM} merge. 
A lightweight control loop runs every few hundred milliseconds and adjusts $r$ and $\tau_{\text{pre}}$ using real-time feedback to improve the KVCache link bandwidth $B_{\text{kv}}$ toward its target $B_{\text{kv}}^{\star}$, reduce the prefill P95 wait (a proxy for TTFT), and lower the fraction of decode time stalled by RAG.
When $u_{\text{kv}}<u_{\text{kv}}^{\star}$ the scheduler increases $r$ or shortens $\tau_{\text{pre}}$ in order to accelerate prefill, while rising decode stalls decrease $r$ so that $Q_{\text{dec}}$ occupies more of $N$. 

This design intentionally makes decode as absorption point and gives prefill first class latency protection. Prefill receives earliest deadline first with step aware slack, a reserved batch share $r$, and a short $\tau_{\text{pre}}$, therefore the KV cache is primed quickly and the KV link remains saturated rather than oscillating.

\bibliographystyle{acm}
\bibliography{sample}

\end{document}